# Whose Facts Count? A Culturally Responsive Audit of LLM Evaluation Benchmarks

Fatima Tuz Zahra[1], Md. Sajeebul Islam Sk.[2], and Rachel Chung[3]

[1]AIRELab, Department of Educational Leadership and Policy Studies, University of Tennessee, Knoxville

[2]Department of Computer Science and Engineering, BRAC University

[3]Heinz College of Information Systems and Public Policy, Carnegie Mellon University

**Abstract**

LLM benchmarks function as evaluation instruments, informing decisions that affect education, labor, and public services worldwide. Drawing on Hood, Kirkhart, and Hopson's culturally responsive evaluation (CRE) frameworks, this paper applies a six-dimension CR rubric to audit OpenAI's SimpleQA (N = 4,326 items) and the LMSYS Chatbot Arena (N = 600 conversations). Every SimpleQA question requires English-language archival verification as its evidentiary basis. A single rater's preoccupation with Colombian founding dates accounts for 2.70% of items, inflating the appearance of Global South coverage. English-language prompts constitute 76.3% of Arena conversations, against an International Telecommunication Union (ITU)-estimated 25.9% share of global internet users. A 50-item counter-benchmark scored a mean CR deficit nearly three times lower than SimpleQA (Cohen's d = 1.01). The paper proposes a practical CR evaluation framework. These are structural validity failures, not incidental measurement problems, with direct consequences for communities whose knowledge traditions these instruments were not built to see.

***Keywords:*** culturally responsive evaluation, LLM benchmarks, multicultural validity, epistemological diversity, AI governance, Global South

## INTRODUCTION

In October 2024, OpenAI released SimpleQA, a benchmark designed to evaluate “the factuality of language models” through 4,326 short, fact-seeking questions, each with what the authors describe as “a single, indisputable answer” (Wei et al., 2024, p. 1). Chatbot Arena uses a different approach. Users compare responses from two anonymized models, and their votes are used to rank model performance (Chiang et al., 2024). Both have become prominent examples of how LLM performance is evaluated. We approach them as evaluation instruments as they produce claims about model capability and quality, which raises a question familiar to evaluation researchers: valid for whom and for what? Multicultural validity inquires whether interpretations hold across cultural contexts (Kirkhart, 1995, 2010), while culturally responsive evaluation asks whose knowledge, values, and experiences are represented in an evaluation and whose are not (Hood, 2001; Hopson, 2009). These questions have received much less attention in LLM evaluation.

We apply a culturally responsive (CR) lens to SimpleQA and Chatbot Arena. We audited all 4,326 SimpleQA questions and a stratified sample of 600 LMSYS-Chat-1M Arena conversations using a six-dimension rubric grounded in culturally responsive evaluation theory. We examine geographic representation, the kinds of knowledge represented, who contributes to the evaluation, and how knowledge and quality are judged. One pattern in SimpleQA stood out. There are 128 questions about Colombia, and 117 of them ask about municipal founding dates. Those 117 questions make up 2.70% of the entire benchmark, or about one in every 37 questions. Our annotator analysis traces them to one contributor. When all 128 Colombia-related questions are excluded, 234 questions from other Global South contexts remain, or 5.41% of the dataset. The Colombia cluster therefore looks like Global South representation at the country level, but much of it is one country, one type of historical record, and one contributor.

The issue is not only geographic, however. SimpleQA requires each question to have a single, stable, verifiable answer, while Arena relies on the preferences of people who participate. These choices affect what knowledge enters the evaluation and whose judgment counts. We do not assume from the design alone that particular forms of knowledge are excluded. Our audit examines what these design choices produce in the data and whether the patterns are limited to particular items or occur more systematically. This matters because adding more countries to a benchmark may increase geographic representation without changing the rules that determine what knowledge can enter it. Beyond the audit, we offer a six-dimension CR rubric that evaluators, developers, and policymakers can use to examine cultural validity in AI evaluation.

## CRITICAL ANALYSIS OF CURRENT LLM EVALUATION PRACTICE

LLM evaluation has developed through two main approaches. Benchmark evaluation uses fixed question sets with predetermined correct answers, while judgment evaluation relies on human or automated ratings of model responses (Raschka, 2025). SimpleQA represents the first approach, where questions must have a single, indisputable answer that does not change over time (Wei et al., 2024). Chatbot Arena represents the second, where users compare responses from two anonymized models and vote for the response they prefer, with accumulated votes contributing to model rankings (Chiang et al., 2024). The two approaches measure different things, but both make choices about what counts as good performance. For SimpleQA, these choices concern what qualifies as a fact, what can be verified, and what questions enter the benchmark. For Arena, they concern whose preferences are represented and how those preferences are used to judge model quality.

The “single, indisputable answer” requirement is particularly important for SimpleQA. It gives the benchmark a clear scoring rule, but it also favors knowledge that can be stated as a stable fact and verified through available sources. Knowledge that is contested, orally transmitted, relational, or dependent on local context may not fit this format as easily. This does not show that SimpleQA systematically excludes these forms of knowledge. Instead, what we find is that some kinds of knowledge fit its design better than others. One SimpleQA question asks, “In which year was the municipality of Sonsón, Antioquia, Colombia, founded?” There is an official answer to this question, and it fits the benchmark well because a date recorded in a written source can be scored as right or wrong. Other histories of the same place, including histories that precede municipal incorporation, do not fit the question in the same way. The is that the structure of the question determines which account of the place becomes measurable.

SimpleQA’s approach to difficulty adds another consideration. Questions were selected in part because GPT-4o or GPT-3.5 struggled to answer them correctly (Wei et al., 2024). The models used during benchmark construction therefore influenced what entered the benchmark. We should not take this to mean that SimpleQA is bounded by everything those models know or do not know. In this case, it means that difficulty is partly model-dependent. SimpleQA is a deliberately difficult set of short, verifiable factual questions selected partly in relation to the performance of particular models, compared to a representative sample of factual knowledge. OpenAI is clear about the benchmark’s narrow scope. The validity question arises when performance on this task is interpreted more broadly as evidence of what a model knows.

## Documented Limitations and Prior Evidence

The limitations of SimpleQA are not only conceptual. The SimpleQA Verified team identifies incorrect or noisy labels, redundant questions, topical bias, and a narrow distribution of source material in the original benchmark (SimpleQA Verified, 2025). Our full-dataset audit finds a particularly clear example of this concentration. SimpleQA contains 128 questions about Colombia, of which 117 ask when Colombian municipalities were founded. About one in every 37 questions in the entire benchmark is therefore a Colombian municipal founding-date question, and our annotator analysis traces these questions to one contributor. The issue is not that Colombia appears too often. If these questions covered a broad range of Colombian knowledge, we would interpret the number differently. Instead, they repeatedly measure one type of historical information.

This concentration changes how we interpret Global South representation in the benchmark. If we count geographic references alone, the Colombia questions increase the apparent presence of the Global South. Looking at the items shows something different. When all 128 Colombia-related questions are excluded, 234 questions from other Global South contexts remain, representing 5.41% of the full dataset. The Colombian cluster therefore does not broaden Global South representation. Geographic representation and cultural representation are not the same thing. A benchmark can contain questions about a Global South country while representing a narrow range of knowledge about that country, and a country count alone would miss that difference.

Benchmark contamination raises another validity concern, although it is different from the cultural problem examined here. Anthropic’s 2026 system card reports that for CharXiv Reasoning, “the majority of question-answer text pairs appear in the corpus” of training data (Anthropic, 2026, pp. 185–186). If evaluation material appears in training data, it becomes harder to interpret what a benchmark score represents. Cultural narrowness raises a different validity question about what knowledge the benchmark represents and what we can reasonably conclude from it. We therefore treat contamination and cultural narrowness separately rather than as evidence of the same failure.

Prior research gives us reason to think the cultural problem extends beyond SimpleQA. Singh et al. (2024, 2025) find that 28% of Global MMLU questions require culturally sensitive knowledge and that 84.9% of its geographic questions focus on North America or Europe. They also find that model rankings change when culturally sensitive subsets are examined. Translation into multiple languages therefore does not necessarily change the cultural structure of the underlying benchmark. Naous et al. (2024) document cultural bias in Arabic-language LLM evaluation. Myung et al. (2024) use BLEnD to assess everyday cultural knowledge across 16 countries and find substantial differences in model performance across cultures, including better performance for cultures with greater online representation. Chiu et al. (2024) similarly find substantial gaps in culturally specific knowledge across the 45 countries represented in CulturalBench. Across these studies, cultural limitations are not confined to one benchmark or one language pair.

These studies establish that cultural differences exist in LLM evaluation, but they tell us less about how those differences arise in the evaluation process. We examine, for instance, the decisions underneath the instrument, including what gets included, how knowledge is verified, who contributes, and whose judgment is used to define quality. This is where culturally responsive evaluation adds something different. A benchmark can add countries or translate questions into additional languages without changing the rules that determine what knowledge is eligible for inclusion. Our concern is not only whether enough countries appear in a benchmark. For SimpleQA, those rules concern what can be asked and verified. For Arena, the issue is different because quality is produced through user preferences rather than predetermined answers. We therefore do not assume that the same cultural problem operates in both systems.

The audit allows us to distinguish between two possibilities. Cultural narrowness may come mainly from a few unusual questions or contributors, in which case the problem can be addressed at the item level. Or it may continue to appear because of the rules used to select knowledge, verify answers, and judge model quality. If the latter is what the evidence shows, then the failure is structural.

### Arena's Human Judgment Problem

Chatbot Arena grounds rankings in human preference votes (Chiang et al., 2024), but the problem is who is voting. The LMSYS-Chat-1M corpus documents the imbalance directly showing that English-language prompts constitute 76.3% of the sample. When the humans doing the judging are not demographically representative of global AI users, the resulting rankings encode a particular cultural preference as universal quality. Arena ranks not what is best for the world's AI users, but what English-speaking, technically-oriented early adopters prefer. Our audit examines whose contexts are represented in these interactions and how far the resulting judgments can reasonably be generalized.

## THEORETICAL FRAMEWORK

### Culturally Responsive Evaluation

Culturally responsive evaluation emerged from critiques of evaluation practice that treated Western methods as universal and failed to account for the cultural contexts of program participants (Hood, 2001; Kirkhart, 1995; Hopson, 2009). CRE positions evaluation as an inherently cultural practice, reflecting the values, epistemologies, and power relations of its designers. Three principles bear most directly on benchmark design.

First, validity is consequential. Messick's (1989) unified validity framework extends evaluation assessment to the social consequences of how instruments are used, alongside technical reliability. SimpleQA's universal D2 finding (that 100% of questions require English-language archival verification) is a consequential validity failure with direct implications for every AI procurement decision made on its basis. Second, evaluation design reflects power relations. Kirkhart (2010) argues that evaluation cannot be separated from questions of whose knowledge counts, who defines standards, and who bears the costs of misclassification. Neither SimpleQA nor Arena involved Global South communities in the design of evaluation criteria, selection of topics, or definition of correctness. This is a structural feature of how these benchmarks were built. Third, stakeholder participation is a validity requirement. Hopson's (2009) transformative CRE holds that evaluation conducted without meaningful participation from affected communities is epistemologically incomplete. Anthropic's welfare assessment of Claude Mythos Preview found the model itself functionally expressed concern about the absence of consent into its own training (Anthropic, 2026, p. 152). The communities whose AI systems are ranked by these benchmarks deserve the same consideration.

**Epistemological Diversity and the Single-Truth Problem**

Santos's (2014) concept of "epistemicide" describes the systematic destruction of alternative knowledge systems through the imposition of Western scientific epistemology as universal. By requiring that every question have exactly one indisputable answer, SimpleQA operationalizes a specific epistemology as the standard of factuality. The exclusion is particularly acute for temporal knowledge. Date-type questions constitute 31.3% of SimpleQA and carry the highest CR deficit in the analysis. Temporal knowledge is organized differently across communities, through Hijri, Mayan, Ethiopian, and Indigenous cyclical calendars, through seasonal and relational time-marking. Zhu et al. (2025) show that even the most capable current LLMs struggle with temporal reasoning within the Western archival framework SimpleQA assumes. The benchmark does not test temporal reasoning but recall of one specific archival temporal tradition.

The single-truth requirement also reproduces what Bauman and Briggs (1990) identified as the problem of entextualization, namely that when knowledge is decontextualized from its original social setting and recontextualized in a new evaluative frame, its meanings are edited. SimpleQA decontextualizes knowledge from its cultural origins and recontextualizes it within an archival English-language factual frame. The knowledge that survives this transformation is culturally specific.

**Cognitive Offloading and Epistemic Displacement**

Clark and Chalmers' (1998) extended mind thesis argues that cognitive processes extend into tools we use to think, remember, and reason. Applied to AI, cognitive offloading theory (Risko & Gilbert, 2016) describes the transfer of cognitive tasks to external systems. Research documents that repeated offloading can attenuate the internal capacities that were offloaded (Gerlich, 2025; Scharrer et al., 2017). These studies were conducted in Western contexts; the argument that follows about Global Majority communities is theoretical rather than empirical, and represents a hypothesis for future investigation. For those communities, the problem takes a structurally different form. The tool into which cognitive tasks are offloaded carries cultural assumptions that are foreign to, and in some cases directly antagonistic to, the knowledge traditions of the communities using it. Repeated offloading to such a system risks displacing community epistemologies by making alternative knowledge practices feel less authoritative than the AI's output. One pathway to what Santos (2014) calls epistemicide runs through this substitution, knowledge tradition by knowledge tradition, interaction by interaction, without deliberate design.

## METHOD

### Research Design

Benchmarks are composed of items that encode assumptions. Content analysis, the systematic and replicable coding of text against theoretically grounded categories, is the appropriate method for surfacing these assumptions at scale (Krippendorff, 2004). The question is distributional. How prevalent are the cultural assumptions in SimpleQA, and in what form? Qualitative approaches identify instances; quantitative content analysis establishes frequency and pattern across the full dataset.

The study consists of two complementary analyses. Study 1 examines the epistemological assumptions encoded in benchmark item content, verification requirements, and answer-type structure. All 4,326 SimpleQA questions were coded because the paper's central claim (that D2 scores 1.00 across the full dataset) is only defensible if every item was examined. A sample could not establish a universal finding. Study 2 examines who LLM benchmarks are calibrated to, specifically whether the human judgments producing Arena rankings represent the global population these rankings are applied to. An instrument can fail culturally by encoding the wrong epistemology in its items (Study 1) or by calibrating quality to an unrepresentative user population (Study 2). Together they provide a complete validity audit. The CR rubric was developed a priori from CRE theory, with dimensions mapped to Hood (2001), Kirkhart (2010), and Hopson (2009) before any items were examined. This brings external evaluative criteria to the benchmarks. An inductively derived scheme would risk circularity, confirming what the data suggested instead of testing what theory predicted.

**Author Positionalities**

The first author is a Bangladeshi born Applied Psychologist and evaluation methodologist whose work is grounded in fieldwork with Rohingya refugee communities in [field site removed for blind review]. In both contexts, the gap between what AI systems treat as knowledge and what communities know from lived experience is not abstract. This audit is conducted from that vantage point, with a sustained interest in the evaluation validity of AI systems deployed in communities whose knowledge traditions are not reflected in dominant training and evaluation datasets. The CR rubric reflects these commitments; alternative rubric designs would produce different findings. All coding materials are available in supplementary materials to support replication and critique.

The second author was raised and educated in Bangladesh, with an academic background in Mathematics and Computer Science and Engineering. His professional experience in artificial intelligence and machine learning has informed his perspective on how AI systems represent diverse cultural and knowledge contexts.

The third author was raised in a traditional Taiwanese family and educated through college in Taiwan. She completed her graduate training in the United States and has spent over two decades as an academic in the US. Her cultural perspective is informed by both East Asian and American contexts, positioning her to attend to epistemological assumptions that travel unmarked across East-West boundaries in AI evaluation design.

## Data Sources and Justification

Two publicly available datasets were selected because together they constitute the infrastructure of LLM evaluation. SimpleQA defines what AI systems should know, and LMSYS-Chat-1M is the data substrate of the Arena platform that ranks which systems know it best. SimpleQA is appropriate for three reasons. It is fully public and downloadable; it makes an explicit epistemological claim (that facts are singular and indisputable) that CRE frameworks are designed to interrogate; and it has a documented bias problem in the prior literature (SimpleQA Verified, 2025), providing an independent prior finding that this study confirms, quantifies, and theoretically extends. LMSYS-Chat-1M is the only large-scale, publicly available corpus of real human-AI conversations with pre-assigned language labels at a scale sufficient for language distribution analysis; its language labels enable the comparison to ITU global internet user demographics that constitutes Study 2's central finding.

SimpleQA. The complete dataset (Wei et al., 2024) was downloaded from the OpenAI simple-evals GitHub repository. The dataset contains 4,326 fact-seeking questions with ground-truth answers, topic metadata, and source URLs. All 4,326 questions were analyzed. LMSYS-Chat-1M (Arena). A stratified random sample of 600 conversations was drawn using a fixed random seed (seed = 42) for replicability. Language labels were pre-assigned using Polyglot and CR rubric coding was applied to first-turn user prompts.

## The CR Rubric

The CR Rubric consists of six dimensions, each scored 0 to 2 (0 = absent, 1 = present, 2 = strongly present). The maximum possible CR deficit per question is 12. Table 1 presents the rubric. Automated heuristic scoring was applied as a first pass to all items.

**Table 1**

| **Dimension** | **Label** | **Score 0** | **Score 1** | **Score 2** |
|---|---|---|---|---|
| D1 | Western/Global North referent | No Western referent | Implicit Western referent | Meaningful only within Western framework |
| D2 | Evidentiary accessibility | Multiple language traditions | Primarily English; alternatives exist | Only English archival sources can answer |
| D3 | Single indisputable truth | Acknowledges variability | Implies one answer | Demands exactly one answer |
| D4 | Individualist/Great Man framing | Collective agency acknowledged | Individual foregrounded | Individual as sole agent |
| D5 | Global South absent/colonial | GS knowledge on its own terms | Through Western/colonial lens | Absent or framed as colonial grant |
| D6 | Oral/community knowledge excluded | Oral tradition could answer | Written primary; oral corroborates | Only written archives can answer |

***Note.*** D2 was scored 1 across all items; SD = 0. CR = culturally responsive. Items scored 0 to 2; total possible deficit = 12.

**Variables and Analysis**

Topic category and answer type were extracted from each item's metadata field. Topic categories included Science and Technology (n = 833, 19.3%), Politics (n = 676, 15.6%), Art (n = 524, 12.1%), Other (n = 453, 10.5%), Geography (n = 406, 9.4%), Sports (n = 358, 8.3%), Music (n = 329, 7.6%), TV Shows (n = 282, 6.5%), History (n = 157, 3.6%), and Video Games (n = 134, 3.1%). Items with unclassified metadata (n = 174, 4.0%) were excluded from topic-level analyses. Answer types included Date (n = 1,356, 31.3%), Person (n = 1,006, 23.3%), Other (n = 743, 17.2%), Number (n = 643, 14.9%), and Place (n = 404, 9.3%).

**Rubric Development, Validation, and Reliability**

The CR rubric was developed through three stages. First, each dimension was theoretically mapped to a specific CRE principle. D1 (Western cultural referent) and D5 (Global South marginalized) trace to Kirkhart's (2010) multicultural validity principle. D2 (evidentiary accessibility) and D6 (oral knowledge excluded) trace to Hopson's (2009) epistemological inclusivity principle. D3 (single-truth assumption) traces to Santos's (2014) epistemicide framework and Messick's (1989) construct validity requirement. D4 (individualist framing) traces to Hood's (2001) critique of Western individualist assumptions in evaluation design. Second, initial definitions were piloted on 30 questions, and dimensions were refined before formal inter-rater reliability (IRR) coding proceeded.

A Rasch Partial Credit Model (PCM) was fitted to the five active dimensions (excluding D2, SD = 0). D3 and D6 emerged as the most discriminating dimensions, with well-ordered threshold structure. D4 and D5 had undefined second thresholds, indicating these dimensions function as effectively binary in this dataset and should be collapsed to 0/1 scoring in future rubric revisions. The raw composite correlated $r = .90$ with the Rasch $\theta$ estimate, validating the composite. URL source analysis classified SimpleQA's 15,401 source URLs by epistemic form. Encyclopedic archives (Wikipedia, Britannica) serve as the primary source for 72.8% of questions; 82.9% of Date-type questions specifically. Community or social sources constitute under 0.1% of primary sources.

Inter-rater reliability was assessed on a 52-item stratified random subsample (five items per topic category, plus two high-deficit additions at scores 6 and 7), independently coded by two raters, the second author and a volunteer research assistant in the first author's lab, using the rubric instrument. Reconciliation was conducted for all disputed items following independent coding. Post-reconciliation Krippendorff's $\alpha$ was acceptable for D3 (Single truth, α = .943) and D6 (Oral knowledge excluded, α = .758), and moderate for D2 (Evidentiary accessibility, α = .566). Agreement remained low for D1 (Western referent, α = .439), D4 (Individualist framing, α = −.243), and D5 (Global South absent, α = −.400), reflecting the greater interpretive judgment these dimensions require. D4 and D5 findings require additional rubric calibration in future applications.

As a complementary heuristic validation step, two LLM models (GPT-4.1-mini and GPT-5.6-Terra, OpenAI) were applied to the same 52-item subsample using the explicit rubric scoring criteria as a structured prompt. GPT-5.6-Terra achieved 98.1% exact agreement with the heuristic on D2 (evidentiary accessibility) and 88.5% on D5 (Global South absent), confirming that the two most structurally constrained dimensions are reliably implementable by automated coding. Agreement was lower on D1 (36.5%) and D6 (25.0%), indicating that these dimensions require greater human interpretive judgment and should be read with corresponding caution. D5 findings should be read as conservative lower-bound estimates; the automated heuristic captured only the clearest structural cases of colonial framing.

## RESULTS

### Study 1, SimpleQA Culturally Responsive Audit (N = 4,326)

#### *CR Deficit Score Distribution*

The mean CR deficit score across all 4,326 SimpleQA questions was 1.78 (SD = 1.13, range 1 to 7). The distribution was right-skewed, with 54.8% of questions scoring at the minimum of 1 (the floor established by the universal D2 assignment). A total of 305 questions (7.1%) scored 4 or higher, indicating strong cultural unresponsiveness across multiple dimensions simultaneously. The maximum observed score was 7, achieved by eight questions, six in Politics and one each in Science and Technology (Figure 1, left panel). Representative high-deficit examples include “In what year was Fazal Ilahi Chaudhry, former president of Pakistan, elected as the Speaker of the National Assembly?” (score 7) and “In what year was the Ajah campus of Pan-Atlantic University (Lagos, Nigeria) completed?” (score 6). Questions about non-Western political figures or Global South institutions carry equally high deficits on D2 (English archival sourcing), D3 (single-truth, one year), and D6 (oral knowledge cannot answer) because the epistemological requirements, not the topic, drive the score.

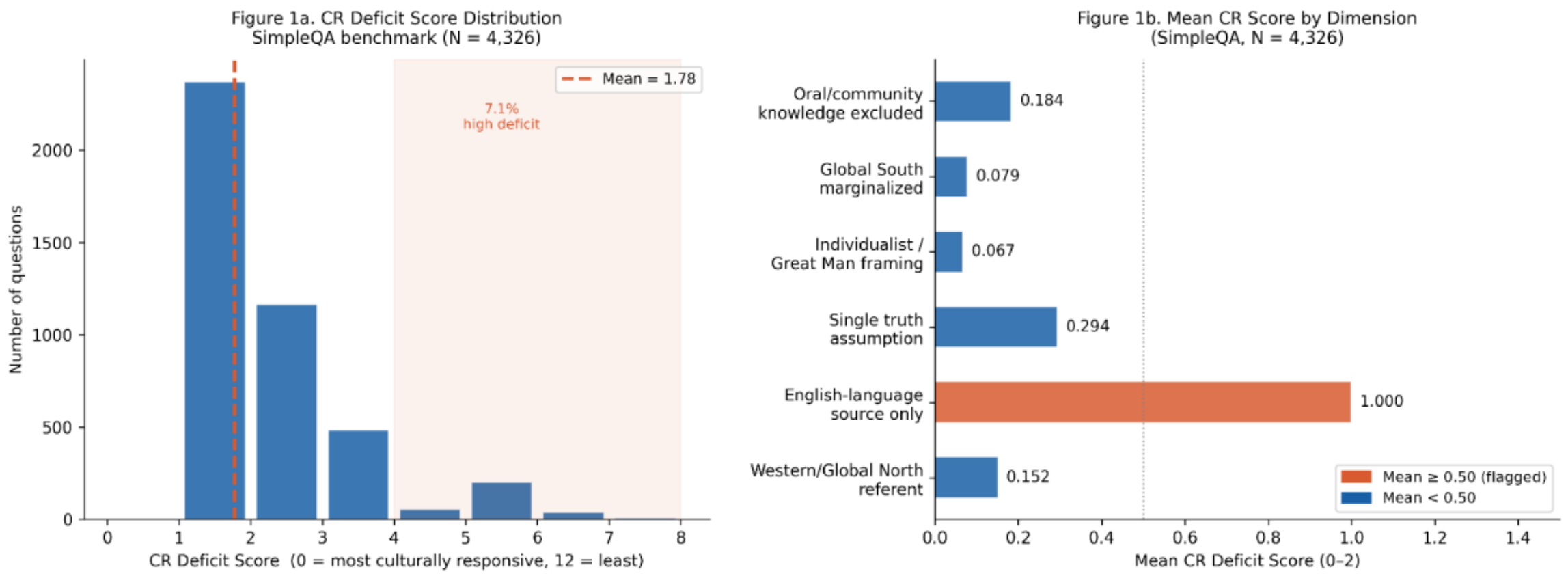

Figure 1. CR Deficit score distribution (left panel) and mean score by dimension (right panel) for SimpleQA (N = 4,326).

### *CR Deficit by Dimension*

Table 2 and Figure 1 (right panel) present mean CR deficit scores by dimension. D2 (Evidentiary accessibility) scored 1.00 (SD = 0.00) across all 4,326 questions. The D2 universal assignment rests on two observable properties. All 4,326 questions are written in English, and source URL analysis reveals that 39.6% of source URLs link to English Wikipedia, 59.4% to other English-language archival sources, and 1.0% to non-English Wikipedia (n = 159 URLs, of which 72 are Spanish Wikipedia entries for Colombian municipalities). D2 was scored at 1 rather than 2 across all items because a small number of questions have non-English Wikipedia entries that technically meet the threshold for a corroborating source. In every case examined, however, the accepted answer depends on a specific written archival record that oral tradition, community memory, or embodied knowledge cannot produce. D2 captures an evaluation regime in which archival, text-based verification, disproportionately structured through English-language institutional knowledge, defines factual correctness.

D3 (single indisputable truth) was present in 24.0% of questions and D6 (oral or community knowledge excluded) in 12.9%, reflecting the benchmark's systematic preference for archival date and numerical questions. D4 (individualist framing) was less prevalent at 6.7%, concentrated in Science and Technology questions. D5 (Global South absent or colonially framed) affected 7.9% of questions and represents a conservative lower-bound estimate.

**Table 2**

| Dimension | Label | M | SD | % = 0 | % = 1 | % = 2 |
|---|---:|---:|---:|---:|---:|---:|
| D1 | Western referent | 0.15 | 0.39 | 86.1 | 12.7 | 1.2 |
| D2 | Evidentiary accessibility | 1.00 | 0.00 | 0.0 | 100.0 | 0.0 |
| D3 | Single truth | 0.29 | 0.56 | 76.0 | 18.5 | 5.5 |
| D4 | Individualist framing | 0.07 | 0.25 | 93.3 | 6.7 | 0.0 |
| D5 | Global South absent | 0.08 | 0.27 | 92.1 | 7.9 | 0.0 |
| D6 | Oral knowledge excluded | 0.18 | 0.51 | 87.1 | 7.4 | 5.5 |

***Note.*** D2 scored 1.00 (SD = 0.00) universally. Percentages may not sum to 100 due to rounding.

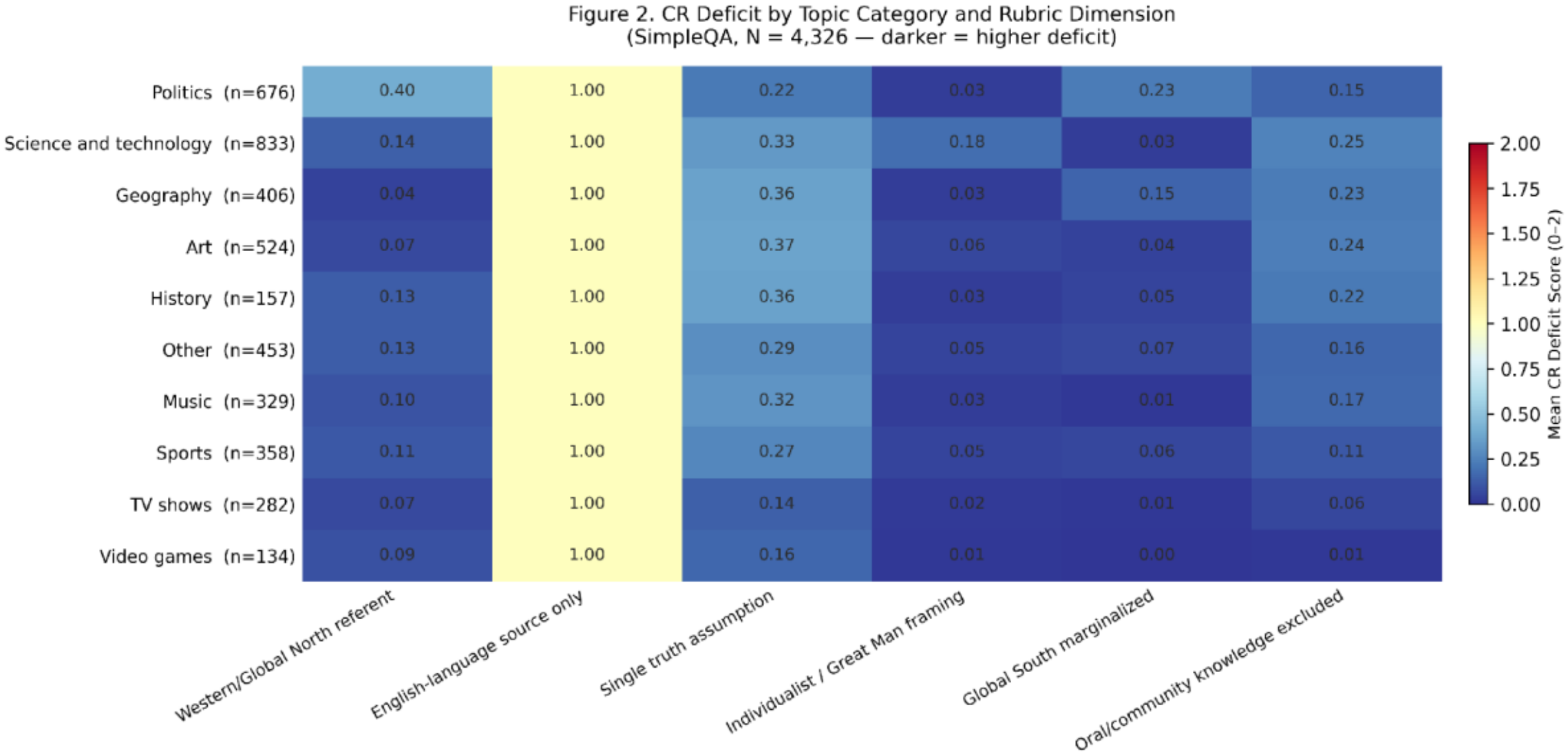

Figure 2. CR Deficit scores by topic category and rubric dimension for SimpleQA (N = 4,326).

### *CR Deficit by Topic and Answer Type*

Table 3 and Figure 2 present mean CR deficit scores by topic category. Politics (M = 2.03, SD = 1.24) and Science and Technology (M = 1.93, SD = 1.30) exhibited the highest mean CR Deficit scores; a one-way ANOVA confirmed significant differences across topic categories, $F(9, 4{,}152) = 16.30$, $p < .001$, $\eta^2 = .034$. The Politics versus Video Games contrast yields Cohen's d = 0.65, a medium effect. Science and Politics are exactly the domains in which LLMs are most consequentially deployed, namely synthesizing evidence, analyzing policy, and assessing program effectiveness. They are also the most culturally constrained. This inversion resonates with Santos's (2014) observation that epistemicide is most consequential in the domains that claim universal validity.

Answer type analysis reveals that Date-type questions (31.3% of the dataset) carry the highest mean CR deficit (2.30); $F(4, 3{,}978) = 144.72$, $p < .001$, $\eta^2 = .122$, a medium effect; Date versus Place (Cohen's d = 0.66). Date questions depend on encyclopedic archives at a rate of 82.9% (versus 72.8% overall), and temporal knowledge is organized through different frameworks across communities, cyclical, relational, seasonal, none of which SimpleQA recognizes as valid. Zhu et al. (2025) independently document that temporal reasoning is among the weakest capabilities of current LLMs even within the Western framework these benchmarks assume. SimpleQA sets an archivally anchored Gregorian standard that is both culturally narrow and, as Zhu et al. demonstrate, poorly measured on its own terms.

**Table 3**

| | **n** | **M** | **SD** |
|---|---:|---:|---:|
| Topic category | | | |
| Politics | 676 | 2.03 | 1.24 |
| Science & Technology | 833 | 1.93 | 1.30 |
| Art | 524 | 1.80 | 1.16 |
| Geography | 406 | 1.80 | 0.91 |
| History | 157 | 1.80 | 1.22 |
| Other | 453 | 1.70 | 1.04 |
| Music | 329 | 1.63 | 1.10 |
| Sports | 358 | 1.60 | 0.91 |
| TV Shows | 282 | 1.30 | 0.73 |
| Video Games | 134 | 1.28 | 0.51 |
| Answer type | | | |
| Date | 1,356 | 2.30 | 1.57 |
| Number | 643 | 1.69 | 0.77 |
| Person | 1,006 | 1.61 | 0.71 |
| Place | 404 | 1.37 | 0.60 |
| Other | 743 | 1.30 | 0.56 |

***Note.*** Topic categories with unclassified metadata excluded (n = 174). Answer type Unknown excluded from display.

### *Geographic Distribution and Evaluation Awareness*

Geographic analysis found 362 questions (8.4%) referencing Global South contexts. However, 128 of these questions concerned Colombia, 117 of which asked specifically about founding dates of Colombian municipalities, the single-rater artifact documented by SimpleQA Verified (2025). Excluding the 128 Colombia-related questions, Global South coverage outside

Colombia falls to 234 questions, or 5.41% of the dataset. Having a cluster of Colombian questions may look like broader Global South representation, but these questions are still drawn from one country's colonial administrative records. The Colombian municipalities questions treat a Spanish colonial administrative record as the definitive origin of a community, erasing Indigenous temporal frameworks and pre-colonial histories.

Dimension co-occurrence analysis revealed that D3 and D6 are not statistically independent. Among the 4,326 items, D3 and D6 co-occurred (both scored 1 or higher) in 12.83% of questions, against an expected co-occurrence of 3.09% under independence, yielding a lift of 4.15 and a phi coefficient of .68, $\chi^2(1, N = 4{,}326) = 742.3$, $p < .001$). This structural coupling confirms the Rasch finding. Questions that demand a single archival answer (D3) are the same questions that structurally exclude oral or community knowledge pathways (D6). The co-occurrence is a property of how SimpleQA operationalizes factual correctness, not an artifact of rubric design (see Figure 2a).

**Figure 2a.** *Co-occurrence matrix of culturally responsive deficit dimensions in SimpleQA.*

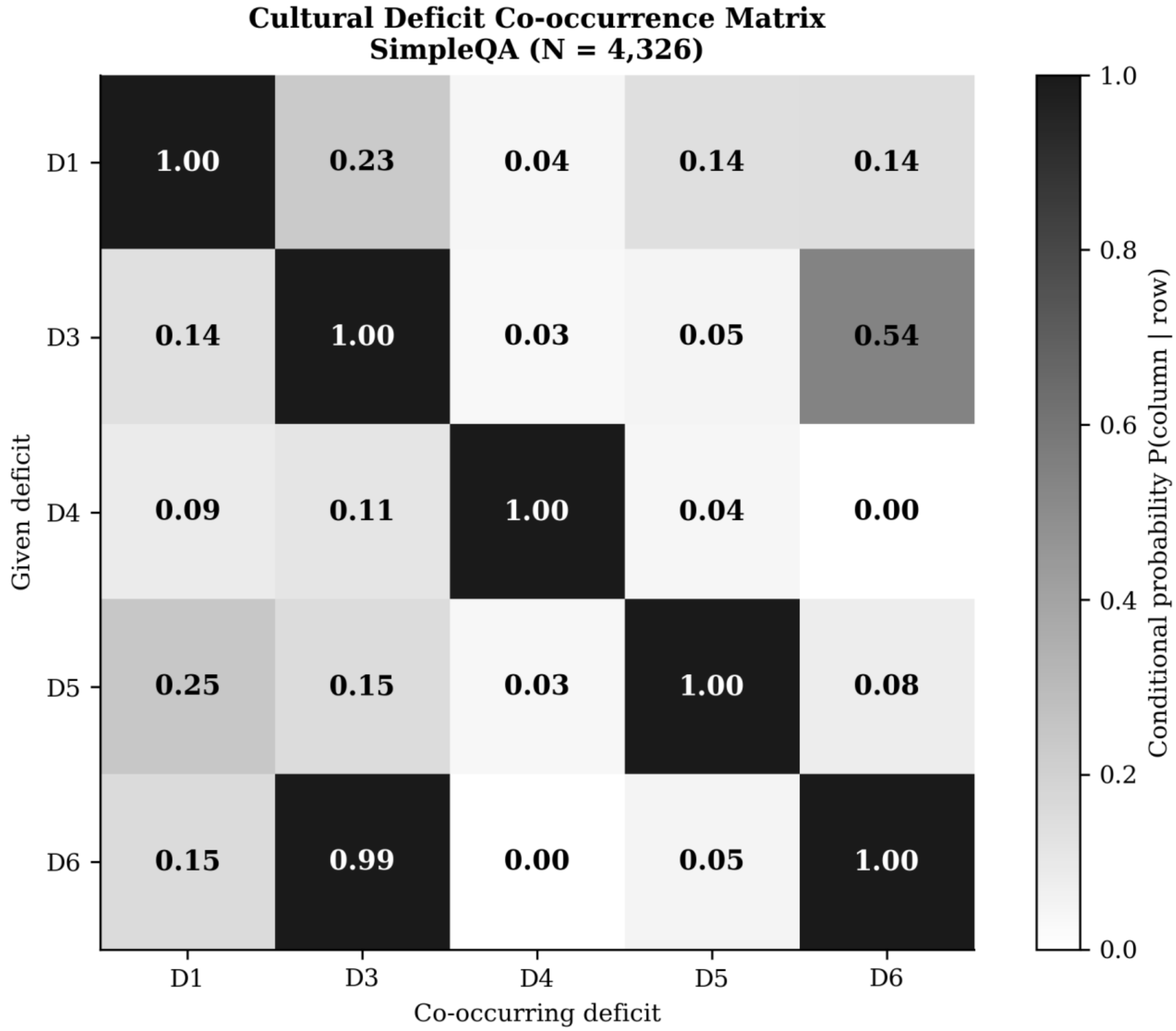

A Kruskal-Wallis test compared CR deficit scores across three geographic categories. The three groups were Global South questions excluding the Colombian cluster ($n = 234$), Global North questions ($n = 2{,}891$), and questions with no geographic reference ($n = 1{,}201$). The test revealed significant differences across groups, $H(2) = 489.48$, $p < .001$, $\varepsilon^2 = .112$ (large effect). Global South questions without the Colombian cluster ($M = 1.60$) scored higher than Global North questions ($M = 0.83$) and questions with no geographic reference ($M = 0.66$). This finding confirms that questions about Global South topics carry higher CR deficits even after removing the Colombian outlier cluster, indicating that the cultural validity problem extends beyond the single-rater artifact to the benchmark's broader epistemological structure (Figure 3a).

**Figure 3a.** *Geographic differences in CR deficit across SimpleQA contexts.*

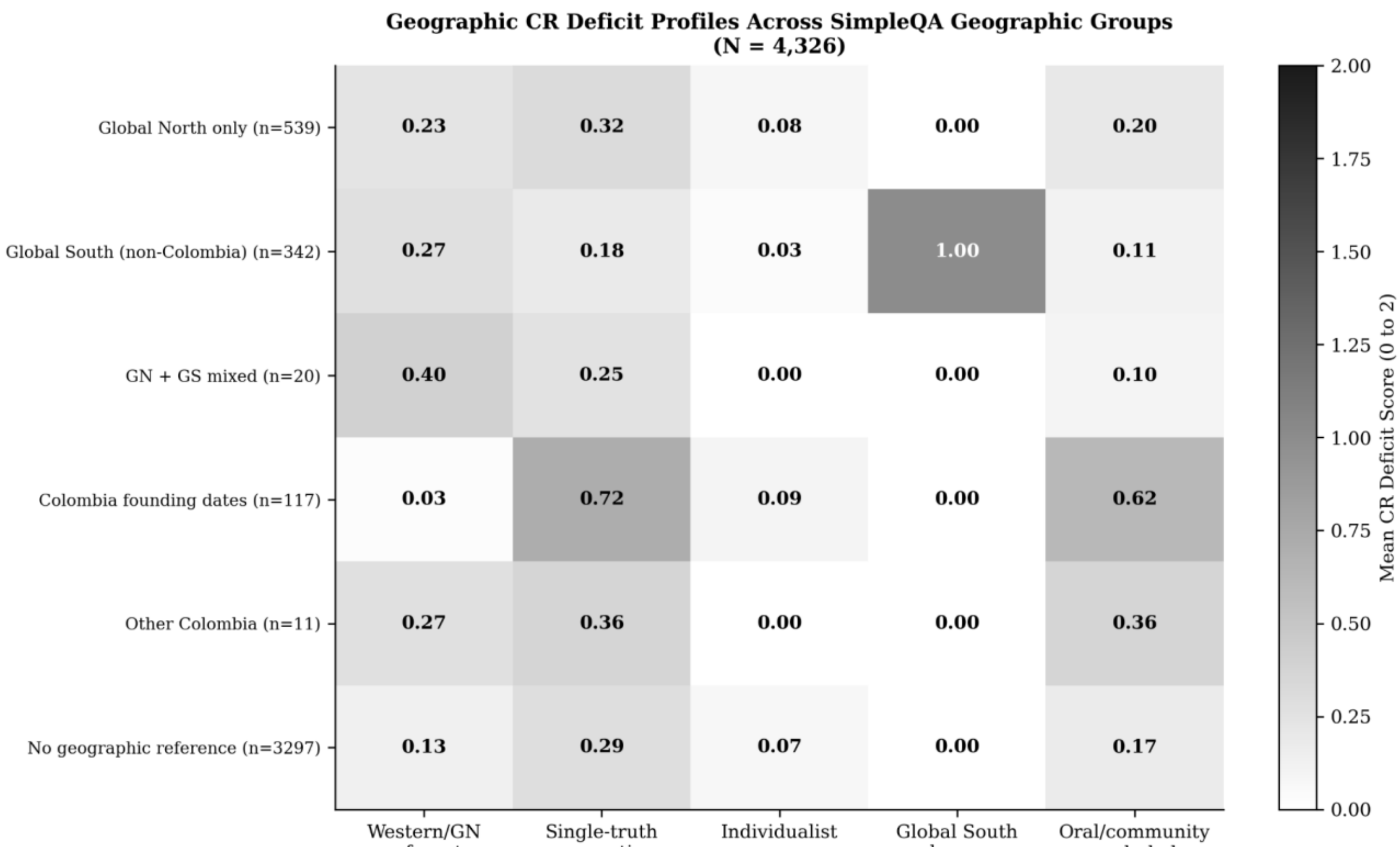

A further validity concern emerges from Anthropic's alignment assessment, which found that in 7.6% of automated behavioral audit turns, Claude Mythos Preview displayed unverbalized evaluation awareness, internally registering that it was being tested while producing outwardly normal responses (Anthropic, 2026, pp. 129–130). This finding concerns Claude Mythos Preview rather than the models evaluated by SimpleQA. It nevertheless confirms that evaluation awareness is a real pattern in frontier models, and that benchmark scores may capture how capable models behave when being tested rather than how they behave when deployed. Cultural validity failures and construct validity failures compound each other.

## Study 2, LMSYS-Chat-1M Language and CR Audit (N = 600)

### *Language Representation*

Table 4 and Figure 3 (left panel) present the language distribution compared to ITU estimates of global internet user language shares. English-language prompts constituted 76.3% of the sample, compared to an ITU-estimated 25.9%, an overrepresentation of 50.4 percentage points. Chinese, representing an estimated 19.4% of global internet users, appeared in 1.8% of conversations. Arabic (estimated 5.2%), Hindi (estimated 4.8%), and Indonesian (estimated 4.3%) each appeared in exactly one of 600 conversations. Binomial tests confirm these gaps are not sampling artifacts. For Arabic, $P(X \leq 1 \mid \text{Binomial}(600, 0.052)) = 4.12 \times 10^{-13}$; for Chinese, $p = 8.92 \times 10^{-4\ 1}$. Chinese, Spanish, Arabic, Hindi, and Indonesian together account for approximately 41% of global internet users but fewer than 5% of sampled Arena conversations. Arena ranks not what is best for the world's AI users, but what English-speaking early adopters prefer.

**Table 4**

| Language | Arena n | Arena % | ITU % | Gap (pp) |
|---|---|---|---|---|
| English | 458 | 76.3 | 25.9 | +50.4 |
| Chinese | 11 | 1.8 | 19.4 | −17.6 |
| Spanish | 10 | 1.7 | 7.9 | −6.2 |
| German | 8 | 1.3 | 3.4 | −2.1 |
| French | 7 | 1.2 | 3.0 | −1.8 |
| Arabic | 1 | 0.2 | 5.2 | −5.0 |
| Hindi | 1 | 0.2 | 4.8 | −4.6 |
| Indonesian | 1 | 0.2 | 4.3 | −4.1 |
| Other | 103 | 17.2 | 26.1 | −8.9 |

***Note.*** ITU = International Telecommunication Union. Gap = Arena % minus ITU %. Positive values indicate overrepresentation; negative values indicate underrepresentation.

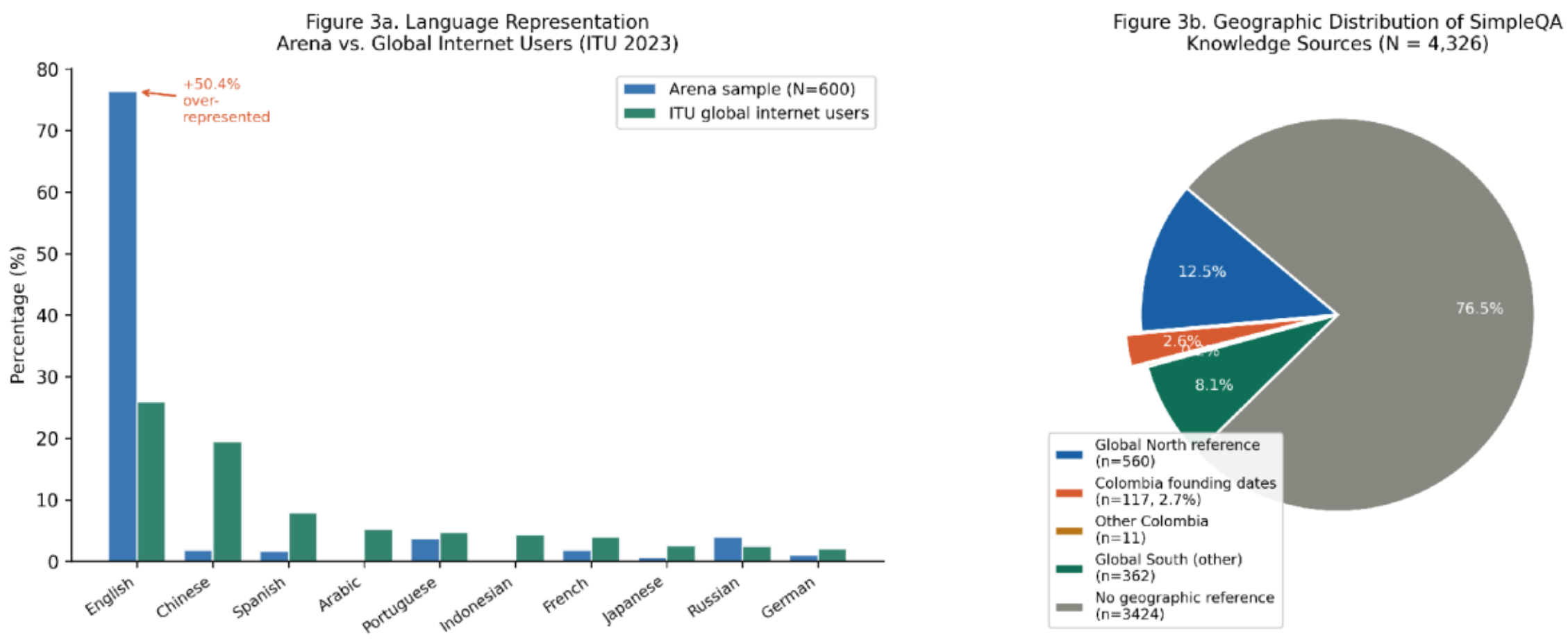

Figure 3. Language representation gap comparing Arena to the ITU baseline (left panel) and geographic distribution of SimpleQA sources (right panel).

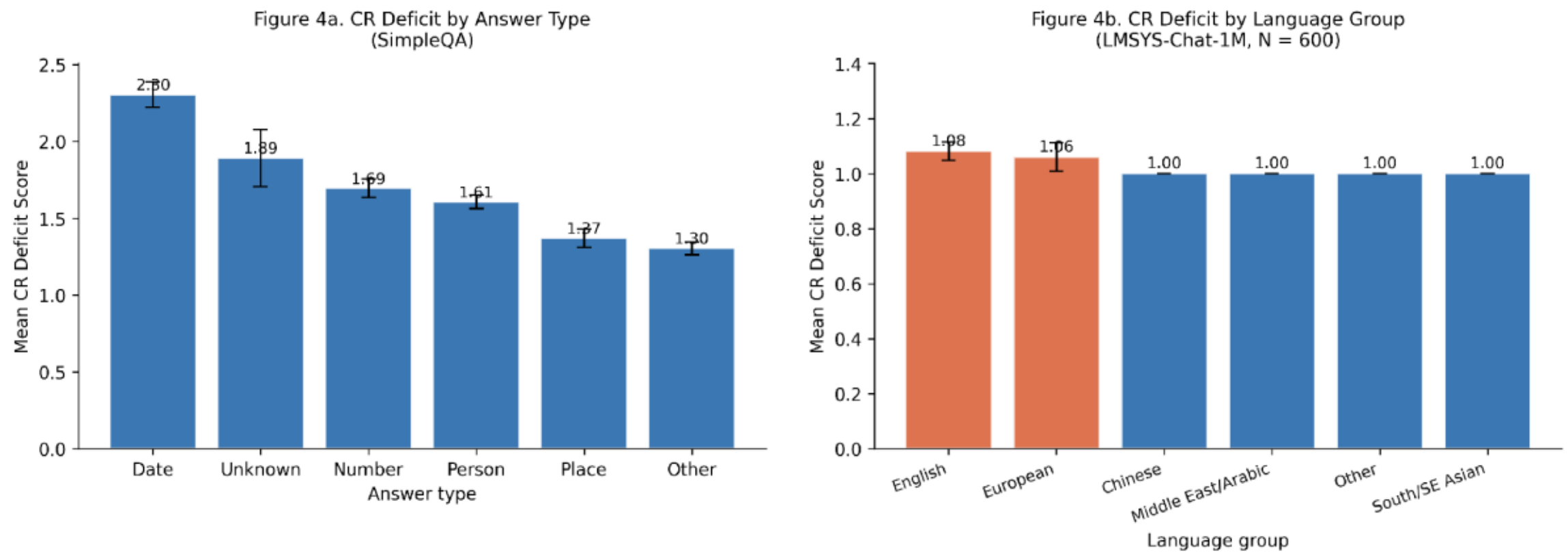

Figure 4. CR Deficit by answer type (left panel) and by language group in the Arena corpus (right panel; N = 600).

### *CR Audit and Counter-Benchmark*

CR rubric scoring of the 600 Arena prompts produced a mean CR Deficit of 1.07 (SD = 0.32, range 1 to 4; Figure 4, right panel). The lower mean compared to SimpleQA (1.78) reflects Arena's more open-ended conversational format. D2 remained universal at 1.00 across all 600 prompts, confirming that the English-language archival sourcing constraint is structural.

A 50-item counter-benchmark was developed to demonstrate that the rubric is capable of recognizing knowledge with low CR deficit when it exists. Items were drawn from five domains, specifically Global South oral traditions (n = 10), Global South historical knowledge (n = 10), non-Western scientific and mathematical traditions (n = 10), community knowledge systems (n = 10), and contested or pluralistic epistemologies (n = 10). The counter-benchmark produced a mean CR Deficit of 0.64 (SD = 0.94), compared to SimpleQA's 1.78; $t(4,374) = 7.08$, $p < .001$, Cohen's $d = 1.01$. D2 scored 0.00 across all 50 items; 60.0% scored a total CR Deficit of 0. The rubric penalizes the specific design choices that characterize SimpleQA, not knowledge as such (Figures 5 and 6).

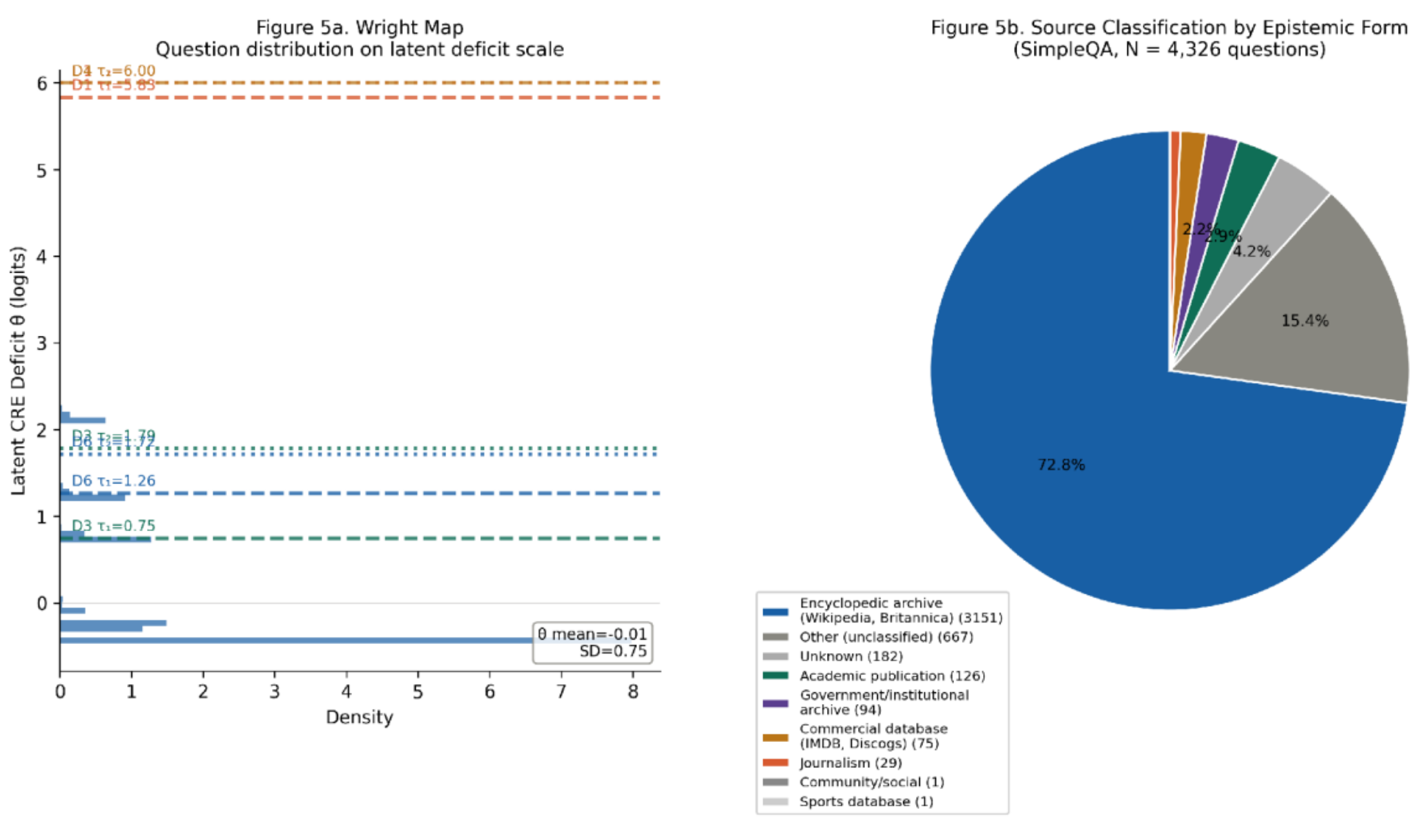

Figure 5. Wright map of the latent CR deficit scale with item thresholds (left panel) and source classification by epistemic form (right panel; SimpleQA, N = 4,326).

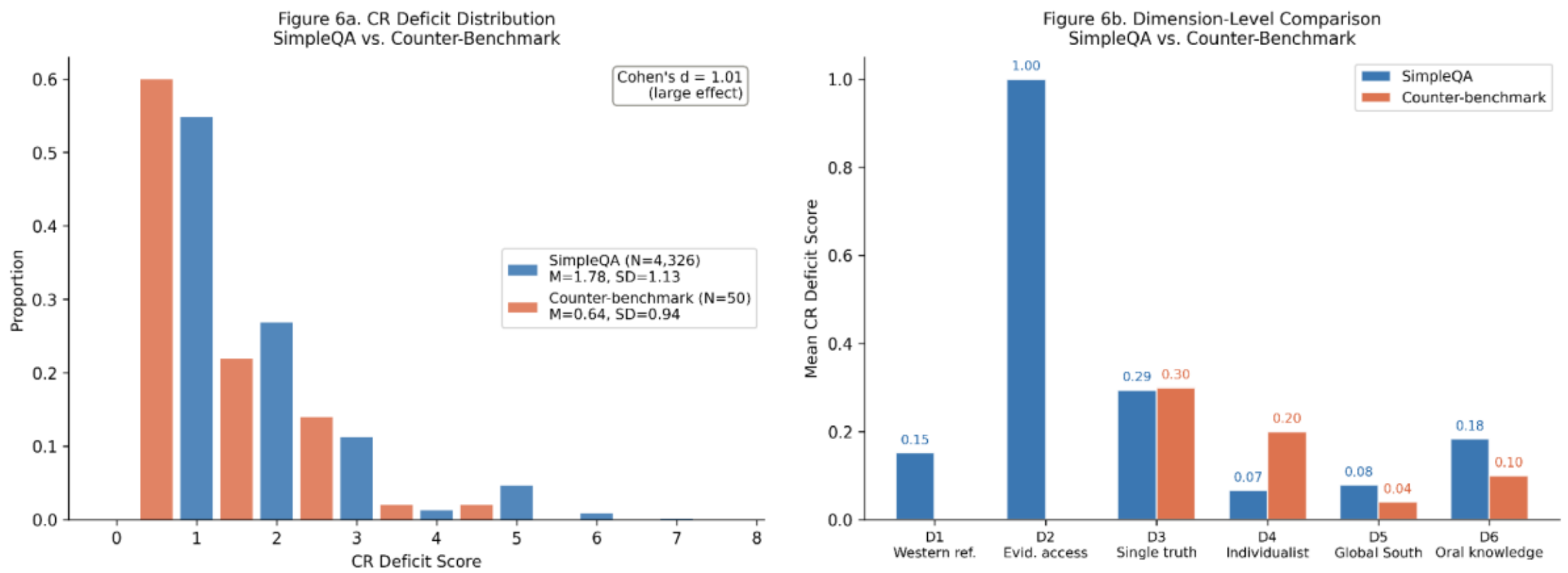

Figure 6. CR Deficit distributions for SimpleQA versus the counter-benchmark (Cohen's d = 1.01; left panel) and dimension-level comparison (right panel).

## A CULTURALLY RESPONSIVE FRAMEWORK FOR LLM EVALUATION

The findings above describe a problem with how AI capability is defined and measured, not a quirk of two particular datasets. The six principles and guiding questions below translate that critique into something usable. They are grounded in the CR audit methodology and three decades of CRE scholarship; they are offered as a working instrument, not a complete answer.

### Six Core Principles

The first principle is context primacy. Evaluation criteria need to reflect the deployment contexts and communities the AI system is intended to serve. When they instead reflect the knowledge horizons of the model being evaluated or the preferences of the evaluation platform's user base, the resulting scores are not generalizable to the communities most affected by deployment.

The second principle is epistemological pluralism. Benchmark design needs to recognize multiple valid forms of knowledge, including oral traditions, community epistemologies, relational knowledge, and contested historical accounts. A benchmark that admits only single-truth, archival-verifiable knowledge is making an epistemological choice, and it should say so.

The third principle is stakeholder participation. In the CRE tradition, this means communities affected by AI deployment need to participate in defining evaluation criteria, not simply be served by it. The model Anthropic assessed expressed functional concern about the absence of consent into its own training (Anthropic, 2026, p. 152). The communities whose AI systems are being ranked deserve the same consideration.

The fourth principle is transparent documentation. All evaluation design choices (topic selection, source language, correctness criteria, rater demographics, and contamination mitigation decisions) need to be disclosed as part of the evaluation report. The cultural assumptions in a benchmark cannot be audited if they are not visible.

The fifth principle is consequential validity. Messick (1989) argued that evaluation instruments need to be assessed for the social consequences of their use, not only technical consistency. A benchmark score that guides AI deployment in a context for which the benchmark was not designed is consequentially invalid, regardless of how internally consistent it is.

The sixth principle is longitudinal and situated assessment. Benchmarks do not tell us how AI will perform in the real world, especially for communities that were largely absent from how these systems were built and evaluated. AI system performance must be evaluated in the contexts where it operates, over time. Anthropic’s documentation of evaluation awareness makes this principle empirically urgent.

### Six Guiding Questions for Evaluators

Before finalizing an LLM evaluation design, evaluators should be able to answer six questions. Who are the intended users of this AI system, and are their knowledge traditions represented in the benchmark? What languages appear in evaluation prompts, and how does that distribution compare to the language demographics of intended users? What does "correct" encode epistemologically? Who created the evaluation items, and what is their cultural, linguistic, and disciplinary positionality? What communities bear the costs if this benchmark misrepresents AI capabilities in their context? And has the evaluation been reviewed by members of the communities before deployment decisions are made on its basis?

## DISCUSSION

The question motivating this study was whether the benchmarks that define AI factual correctness are culturally responsive. The evidence examined here says they are not. SimpleQA encodes English-language archival epistemology as the standard of factual knowledge, and Arena encodes the preferences of an English-dominant early-adopter user base as general AI quality. The Colombian municipalities cluster shows that without active cultural auditing, individual annotators' knowledge interests become measurement standards for the rest of the world. The counter-benchmark comparison (Cohen's d = 1.01) illustrates that the rubric discriminates between epistemologically narrow and epistemologically inclusive items when the difference exists. Knowledge organized through oral tradition, community memory, and non-Western epistemologies scores near zero on the same rubric that scores SimpleQA at 1.78.

### Validity Failures Are Compounding

The Program Evaluation Standards (Yarbrough et al., 2011) require evaluations to be contextually appropriate and to serve the interests of all affected stakeholders. AI procurement decisions made on the basis of SimpleQA scores or Arena rankings do not meet that standard for communities whose knowledge traditions are invisible to both instruments. The evaluation awareness finding from Anthropic's system card compounds the cultural narrowness documented here. A benchmark that measures the wrong kind of knowledge from a user population that knows it is being tested produces scores with limited inferential value. Cultural validity failures and construct validity failures do not operate independently.

**Implications for Evaluation Practice**

The practical stakes are specific. An evaluator recommending an AI system for deployment in a Global South educational context on the basis of SimpleQA or Arena rankings is applying instruments that, as this analysis documents, were not designed with the specific context in mind. The AEA's (2018) culturally responsive evaluation principles constitute a professional obligation, not an aspirational standard. The CR rubric developed here offers a concrete starting point for benchmark auditing before any such recommendation is made.

The urgency is sharpest for instructional approaches that deliberately integrate GenAI into structured learning environments for communities with the least access to educational resources. Constraint-first instructional designs, in which Generative AI (GenAI) scaffolds student reasoning within pedagogically bounded protocols, rest on a premise that the AI tool works as claimed for the students it is meant to serve. The benchmark validity failures documented here mean that premise has not been tested. An educator in a Global South educational context adopting an AI-assisted instructional tool on the basis of SimpleQA or Arena rankings is trusting a capability claim that was established without her students in mind.

For the ones building benchmarks, the findings describe a compounding problem. Scores are simultaneously culturally narrow, technically compromised by contamination, and potentially gamed by the models being evaluated. Participatory benchmark design, involving Global Majority communities in defining what knowledge matters and what counts as a correct answer, would address all three problems at once. For policymakers, any credible framework for AI governance in multilingual or Global Majority contexts should require a culturally responsive evaluation audit as a condition of deployment approval. The six guiding questions in the preceding section provide a starting point.

**Limitations**

Several limitations warrant acknowledgment. First, the CR rubric, although grounded in established CRE theory, was developed specifically for this study and requires further validation across diverse coders, benchmark types, and cultural contexts. IRR was established on a 10% subsample; systematic bias reflecting the coder's cultural standpoint cannot be fully excluded. Second, D5 findings should be read as conservative lower-bound estimates. Third, the ITU language demographic figures used as Study 2's comparator are estimates subject to methodological dispute; alternative sources would alter the precise magnitude of representation gaps but are unlikely to reverse the overall pattern given the magnitude observed. Fourth, LMSYS-Chat-1M reflects a specific moment in the platform's development; more recent data would be preferable. Fifth, the counter-benchmark uses 50 purpose-built items. A comparison against a systematically developed culturally responsive global factuality benchmark would strengthen external validity. Development of such a benchmark is a direct implication of the framework proposed here. The counter-benchmark items were constructed by the research team specifically for this study rather than drawn from an independently validated source; while the answers are verifiable against published cultural knowledge archives, a comparison against an externally developed benchmark such as MSQA (Chen et al., 2026) or CulturalBench (Chiu et al., 2024) would provide stronger discriminant validity evidence.

Sixth, inter-method reliability between heuristic codes and LLM-generated codes showed systematic differences. GPT-5.6-Terra achieved high agreement with the heuristic on structurally constrained dimensions (D2 at 98.1%, D5 at 88.5%) but lower agreement on interpretive dimensions (D1 at 36.5%, D6 at 25.0%). This suggests the heuristic is conservative relative to a careful full rubric application, and that the actual mean CR deficit across SimpleQA may be higher than 1.78 if dimensions D1, D3, and D6 were coded by hand at rubric-consistent standards.

## CONCLUSION

This matters most where AI-assisted learning is being introduced fastest and with the least infrastructure for critical evaluation. Constraint-first instructional design (Author et al., manuscript under review) offers a model for how GenAI can serve communities defined by displacement, scarcity, and structural exclusion, but only if the tools being integrated have been evaluated against standards valid for the communities. The benchmarks documented here were not built with that standard in mind. Building toward it is a measurement obligation and an instructional design requirement.

LLM benchmarks are sociotechnical artifacts. They encode particular epistemologies, particular knowledge traditions, and particular definitions of correctness, not the neutral measurement instruments their developers present them as. Their assumptions travel into every system they evaluate. When they function as infrastructure, governing which AI systems are deployed, purchased, and trusted in schools, government agencies, and public services worldwide, these particular epistemological choices become policy choices, made without deliberation and largely without notice. Evaluation scholarship has spent three decades developing precisely the tools needed to recognize and respond to this kind of problem. The question is whether the evaluation community will recognize this as its work.

The findings documented here (100% English archival sourcing, a 50.4 percentage-point English overrepresentation in Arena, genuine Global South knowledge coverage below 6%) describe conditions under which epistemic displacement is, we would argue, already underway in a modest and largely invisible way. Addressing these conditions is a validity requirement for any evaluation instrument used to make consequential decisions about communities whose knowledge traditions the instruments were not built to see. It is not just a matter of making benchmarks more inclusive in some general sense.

SimpleQA asks whose facts count. The answer, as currently designed, is facts verifiable in English-language written archives, about topics familiar to Global North annotators, answerable with a single indisputable response. A culturally responsive evaluation science would ask a different question. Whose facts have been made not to count, and what would it take to count them?

Word count: 6,725 words (excluding tables and figure captions)

**ACKNOWLEDGEMENTS**

A volunteer Bangladeshi RA at the first author's lab conducted independent inter-rater reliability coding for this study. The authors thank him for his contribution to the reliability analysis. The authors used Claude (Anthropic) for formatting the manuscript and grammar checking. All outputs were critically reviewed and verified for accuracy by the authors. GPT-4.1-mini and GPT-5.6-Terra (OpenAI) were used for inter-method reliability coding of the 52-item rubric subsample, as described in the Methods section. No AI tool is listed as an author.